\documentclass[twocolumn]{aastex631}
\usepackage{graphicx}
\usepackage{float}
\usepackage{lineno}

\newcommand{\lsim }{{\lower0.8ex\hbox{$\buildrel <\over\sim$}}}
\newcommand{\gsim }{{\lower0.8ex\hbox{$\buildrel >\over\sim$}}}
\newcommand{\Msun}{\ifmmode {M_{\odot}}\else${M_{\odot}}$\fi}
\newcommand{\Lsun}{\ifmmode {L_{\odot}}\else${L_{\odot}}$\fi}
\newcommand{\Rsun}{\ifmmode {R_{\odot}}\else${R_{\odot}}$\fi}

\shorttitle{Is the FRB Globular Cluster special?}
\shortauthors{Dage et al.}

\begin{document}

\title{Is the M81 Fast Radio Burst Host Globular Cluster Special?}

\correspondingauthor{Kristen Dage}
\email{kcdage@wayne.edu}

\author[0000-0002-8532-4025]{Kristen C. Dage} \thanks{NASA Einstein Fellow}
\affiliation{Wayne State University, Department of Physics \& Astronomy, 666 W Hancock St, Detroit, MI 48201, USA}
\author[0000-0003-2506-6041]{Arash Bahramian}
\affiliation{International Centre for Radio Astronomy Research Curtin University, GPO Box U1987, Perth, WA 6845, Australia}
\author[0000-0002-6437-6176]{Clancy W. James}
\affiliation{International Centre for Radio Astronomy Research Curtin University, GPO Box U1987, Perth, WA 6845, Australia}
\author{Arunav Kundu}
\affiliation{ Department of Physics, Birla Institute of Technology \& Science, Pilani, K K Birla Goa Campus, NH17 B, Zuarinagar, Goa 403726, India}
\author[ 0000-0001-8283-4591]{Katherine L. Rhode}
\affiliation{Indiana University Department of Astronomy, 727 East Third Street, Bloomington, IN 47405, USA }
\author[0000-0002-1468-9668]{Jay Strader}
\affiliation{Center for Data Intensive and Time Domain Astronomy, Department of Physics and Astronomy, Michigan State University, East Lansing MI, USA}

\author{Enrico Vesperini}
\affiliation{Indiana University Department of Astronomy, 727 East Third Street, Bloomington, IN 47405, USA }

\author{Stephen E. Zepf}
\affiliation{Center for Data Intensive and Time Domain Astronomy, Department of Physics and Astronomy, Michigan State University, East Lansing MI, USA}

\begin{abstract}

We use multiband archival HST observations to measure the photometric and structural parameters of the M81 globular cluster that hosts the Fast Radio Burst FRB 20200120E. Our best-fitting King model has an effective radius $r_h = 3.06$ pc with a moderate King model concentration of $c = 53$, and an inferred core radius of 0.81 pc. 
We revisit the exact astrometric location of the FRB within the cluster, and find that FRB 20200120E is located 1.92 pc
from the center, but within the projected half-light radius. We estimate the relative encounter rate of the FRB host, along with the corresponding rates of 210 other globular clusters in M81, and compare these values with the encounter rates of Galactic globular clusters. The FRB resides in a globular cluster with an encounter rate that is moderately higher than the median stellar encounter rate in our two comparison samples.  While the estimated encounter rate of the FRB host cluster (e.g., $\sim50\%$  of a cluster like 47 Tuc) is sufficient to allow the possibility that the FRB formed dynamically, our results do not place strong constraints on this scenario due to the limitations of the available HST data and the possible systematic uncertainties and selection effects in the comparison data.
\end{abstract}


\section{Introduction} \label{sec:intro}

Fast radio bursts (FRBs) are millisecond-duration radio transient events of unknown origin \citep{Lorimer2007,Thornton2013}. Ever since the localisation of the first repeating fast radio burst, FRB~20121102A, to a dwarf star-forming galaxy \citep{Spitler2016,VLAlocalisation}, young magnetars (strongly magnetized neutron stars that formed less than a few decades ago from Type II supernovae) have been hypothesised as the progenitors of FRBs\citep{Metzger2017magnetar}. This view is backed up by the association of FRB~20121102A with a persistent radio source \citep[PRS;][]{VLAlocalisation} and by that source's extreme magneto-ionic properties \citep{Michilli2018_121102}. On the other hand, the idea that FRBs have a common origin has been challenged by further localisations of FRBs \citep[e.g.][]{Bannister2019}, which reveal that many come from galaxies with lower star-formation rates. The observed properties of the FRBs, such as the radial offset distributions, are inconsistent with the corresponding properties of most other classes of astrophysical transients \citep{Bhandari+22,Gordon2023}. Furthermore, while at least one other FRB (FRB~20190520B) appears similar to 20121102A in  terms of its repetition rate, host galaxy, association with a persistent radio source, and magneto-ionic properties \citep{Niu2022}, some FRBs have shown no evidence of repetition despite significant follow-up campaigns \citep{James2020a_followup,Lee-Waddell_2023,2023CHIEMFarSidelobes}. Still other repeating FRBs show significant offsets from star-forming activity in their host galaxies \citep{MarcoteRepeaterLocalisation2020,Tendulkar2021NoSF}. Overall these findings motivate the consideration of alternative progenitor scenarios. In particular, the pre-merger orbital interactions \citep{Wang2016Inspiral}, merger \citep{Totani2013}, and/or post-merger collapse \citep{FalckeRezollaBlitzar} of old stellar remnants such as neutron stars and white dwarfs have long been proposed as FRB progenitor pathways, though these tend to favour once-off FRBs, or those that repeat for only a very short duration.

The repeating FRB 20200120E was localized to a specific host system --- a globular cluster (GC) in the nearby spiral galaxy M81 \citep{2022Natur.602..585K}, which is approximately 3.6 Mpc away. Globular clusters are extremely old  ($\sim$10-13 Gyr) stellar systems, and provide valuable constraints on possible progenitors to the FRB, including limiting the possibility of a magnetar origin \citep{2022Natur.602..585K}. Another potential theory
is an origin from a hyperaccreting X-ray binary \citep{2022ApJ...937....5S}. Although the FRB source
 was not detected in the X-ray by an off-axis archival \textit{Chandra} observation, nor in additional follow-up X-ray observations, \citep{2022Natur.602..585K, 2023arXiv230810930P}, hyperaccreting X-ray binaries in globular clusters are demonstrated to show orders of magnitude X-ray variability on the scale of hours \citep[and references therein]{2020MNRAS.497..596D}.
Young neutron stars formed as a result of the collapse or merger of white dwarfs present another plausible scenario \citep{Kremer21}. Such mergers would be the result of dynamical interactions in the dense environment of  GCs and are expected to occur mainly in clusters at the time of core collapse or in the post-core collapse phase \citep{Kremerwd, Kremerm87}; white dwarf mergers may also explain the origin of young pulsars and single millisecond pulsars observed in Galactic globular clusters \citep{Kremerwd, 2023ApJ...953..141Y}. 

M81's globular cluster system has been extensively studied with photometry and spectroscopy in the optical and NIR \citep{1995AJ....109.1055P, 1995AJ....110..620P,2007PASP..119.1085M, 2011AJ....142..183N, Pan2022, 2022MNRAS.516.1320C}.  The X-ray source population associated with the globular clusters has also been well characterized \citep{2023ApJ...947...31H}.

Globular clusters hosted by galaxies beyond the Local Group often appear as point sources in ground-based images; in such cases, globular clusters can only be studied via their integrated light.  On the other hand, HST is able to at least partially resolve globular clusters in distant galaxies  \citep[see][among many others]{Whitmore1993,1999ApJ...513..733K, 2005ApJ...634.1002J, Sivakoff07,Strader2011,2012ApJ...759..126P}.  The half-light radii of the clusters can be measured from HST images using routines such as baolab \citep{Larsen99} and assuming that the cluster light profile follows a King model.  This improvement in resolution means that globular clusters are more accurately identified in HST studies than in ground-based data.

The ability to determine cluster size also impacts the estimates of stellar encounter rates in globular clusters. The stellar encounter rate ($\Gamma$) in a cluster is directly linked with the population of close binaries and compact objects in these dense stellar systems \citep{Pooley03, Heinke03d, Bahramian13} and depends on cluster properties as $\Gamma\propto\int\rho^2/\sigma~dV$, where $\rho$ is cluster stellar density, $\sigma$ is velocity dispersion of stars in the cluster and $\Gamma$ is estimated over the volume of the cluster $V$ \citep[e.g.,][]{Hills76, Verbunt87}. The stellar distributions of the large majority of globular clusters are well-described by King models \citep{King62,King66}. In these clusters the encounter rate is dominated by the contribution from the cluster core, leading to an approximation of $\Gamma$ as $\propto \rho_c^{3/2} r_c^{2}$, where $\rho_c$ is the central density and $r_c$ is cluster core radius. Given the large distances of extragalactic globular clusters and thus their small angular sizes, accurately measuring quantities such as the core radius is typically challenging.  This leads to approximations of $\Gamma$ through a proxy such as $\Gamma_h \propto M^{3/2} r_h^{-5/2}$, where $M$ is cluster mass and $r_h$ is the cluster half-light radius, though this proxy measurement is much less sensitive to the encounter rate than the core-based measurements \citep{Sivakoff07}. 

\begin{figure*}
    \centering
    \includegraphics[width=\textwidth]{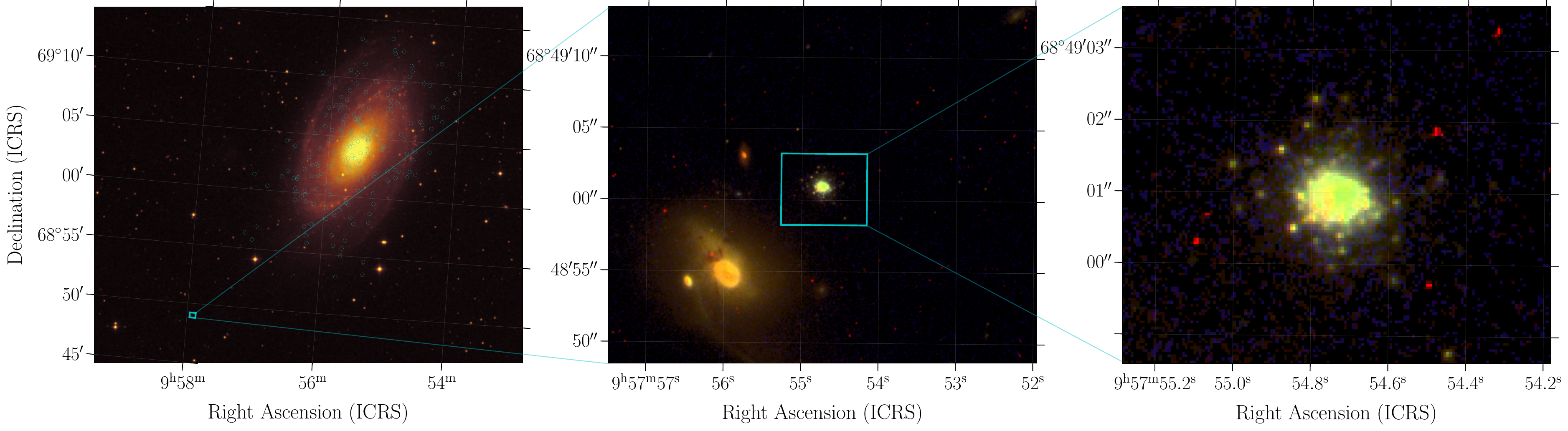}
    \caption{\textit{Left}: Digital Sky Survey (DSS) R band image of M81. Positions of globular clusters from NH11 considered in this work are shown by cyan circles. The cyan rectangle indicates the area around the FRB GC, plotted in the middle and right panels. \textit{Middle} and \textit{right}: A composite color image of the vicinity of the FRB GC (middle) and a zoom-in on the cluster (right) based on HST images in F814W (red), F606W (green), and F435W (blue).}
    \label{fig:HST}
\end{figure*}

The globular cluster that hosts FRB20200120E was labeled with ID number 30244 by \cite{1995AJ....109.1055P} and \cite{1995AJ....110..620P} in their ground-based imaging and spectroscopic studies.  \cite{1995AJ....109.1055P} measured the apparent magnitude and colors for the globular cluster  of V=19.76, B$-$V=0.77, and V$-$R=0.47. We refer to [PR95] 30244 as FRB GC in this work. \citet{1995AJ....109.1055P} acquired spectroscopy of FRB GC (albeit with what they characterize as ``poor signal-to-noise'') to confirm its association with M81, as well as to estimate a highly uncertain metallicity of [Fe/H]=$-1.76\pm1.78$ for the cluster. The FRB GC is located roughly 19.6$'$ from the center of M81, which translates to $\approx$20.5 kpc at the relevant distance. \cite{Pan2022} used multiwavelength archival data and other information from the literature to assemble an accurate list of the globular clusters in M81 and included this object in their catalog. Based on an initial assessment of these various measurements, the FRB GC does not seem to be unusual, despite playing host to an extremely mysterious and energetic radio signal. However, its half-light radius and encounter rate may shed further light on the nature of the cluster and the possible physical cause of the FRB emission. In their paper presenting the discovery of FRB20200120E, \cite{2022Natur.602..585K} combined broadband ugriz photometry from the Sloan Digital Sky Survey (SDSS) of the FRB GC with a stellar population model and a few assumptions (e.g., a model for the star formation rate of the cluster since its formation) to come up with reasonable estimates of several fundamental properties of  the cluster -- e.g., metallicity, velocity dispersion, mass, and effective radius.  They estimated [Fe/H] of -1.83, a cluster stellar mass of $\log(M/\Msun) = 5.77$, a velocity dispersion of 22 km/sec, an effective radius of 3.7pc, and an age of 9.1 Gyr.
    
In this work, we use archival HST images of the FRB GC to measure the optical photometric and structural properties of the cluster and compare them to work from \citealt{2011AJ....142..183N} (hereafter NH11). Our objective is to examine the properties of this cluster relative to those of the other clusters in M81 and the Milky Way and look for clues that might help reveal why this cluster hosts an FRB. The data and analysis methods are described in Section \ref{sec:data}, the results are presented in Section \ref{sec:comparison}, and we summarize our conclusions in Section \ref{sec:summary}.

\section{Analysis \& Results} \label{sec:data}
The M81 FRB GC was observed by the Hubble Space Telescope's Wide Field Camera 3 (Program 16664, PI: Tendulkar) on 2022-04-02 (Orbit 1: 1651 seconds F606W \& F814W, 1810 seconds F438W), 2022-11-15 (Orbit 2: 1639 seconds F606W \& F814W, 1801 seconds F438W), and 2023-02-22 (Orbit 3: 2708 seconds, F606W only) with a 3-point dither pattern.  

\subsection{HST analysis \& cluster half-light radius}

The first two orbits of program 16664 suffered a guide-star failure and guided on gyros only. As a result, only a subset of exposures from this  visit was usable. We examined the calibrated, flat fielded CTE-corrected individual exposure (FLC) files and retained only those images in which the FRB GC was detected. For F438W, these were iem701lxq, iem701lrq from Orbit 1 and iem751ibq and iem751i4q from Orbit 2 (2622 seconds total). For F606W, the usable images were iem701lvq from Orbit 1,  iem751i9q from Orbit 2, and iem752u0q, iem752u1q, iem752u2q, iem752u4q from Orbit 3 (4020 seconds total), and for F814W iem701ltq in Orbit 1,  iem751i7q in Orbit 2 (1312 seconds total). The HST data can be found in MAST at \dataset[10.17909/vysd-m633]{http://dx.doi.org/10.17909/vysd-m633}.

We manually redrizzled the images using the DrizzlePac software (\citealt{2021AAS...23821602H}, stwcs Version 1.7.2, photutils Version  1.7.0), aligning the frames with \textsc{tweakreg} and verifying the shifts manually with \textsc{IRAF} \citep{1986SPIE..627..733T}. In the F438W images, there were too few bright sources present in the field for the software to identify the offset, so we computed the shift manually with IRAF tasks and updated the headers. We drizzled the images with \textsc{astrodrizzle}, with the `minmed' combine type. A composite-color image of the \emph{HST} data is presented in Figure~\ref{fig:HST}.

To determine the structural parameters of the cluster, we began by constructing an empirical point spread function in $F606W$ using three bright but unsaturated stars (confirmed as stars via Gaia DR3). We then subsampled the point spread function by a factor of 10. We carried out the King model fitting \citep{King62} using this subsampled point spread function and the ishape task in the baolab package 0.94.1 \citep{Larsen99}. We tried fitting radii of both 50 and 60 pixels, finding very similar results in each case. The specific results quoted below are for the 50 pixel case.
We used the the WFC3 pixel scale of $0.0396\arcsec$ per pixel \footnote{\url{https://esahubble.org/about/general/instruments/wfc3/}}, and the 3.6 Mpc distance to M81, resulting in a distance scale of 0.69 parsec per pixel. 

\begin{figure}
\includegraphics[width=0.45\textwidth]{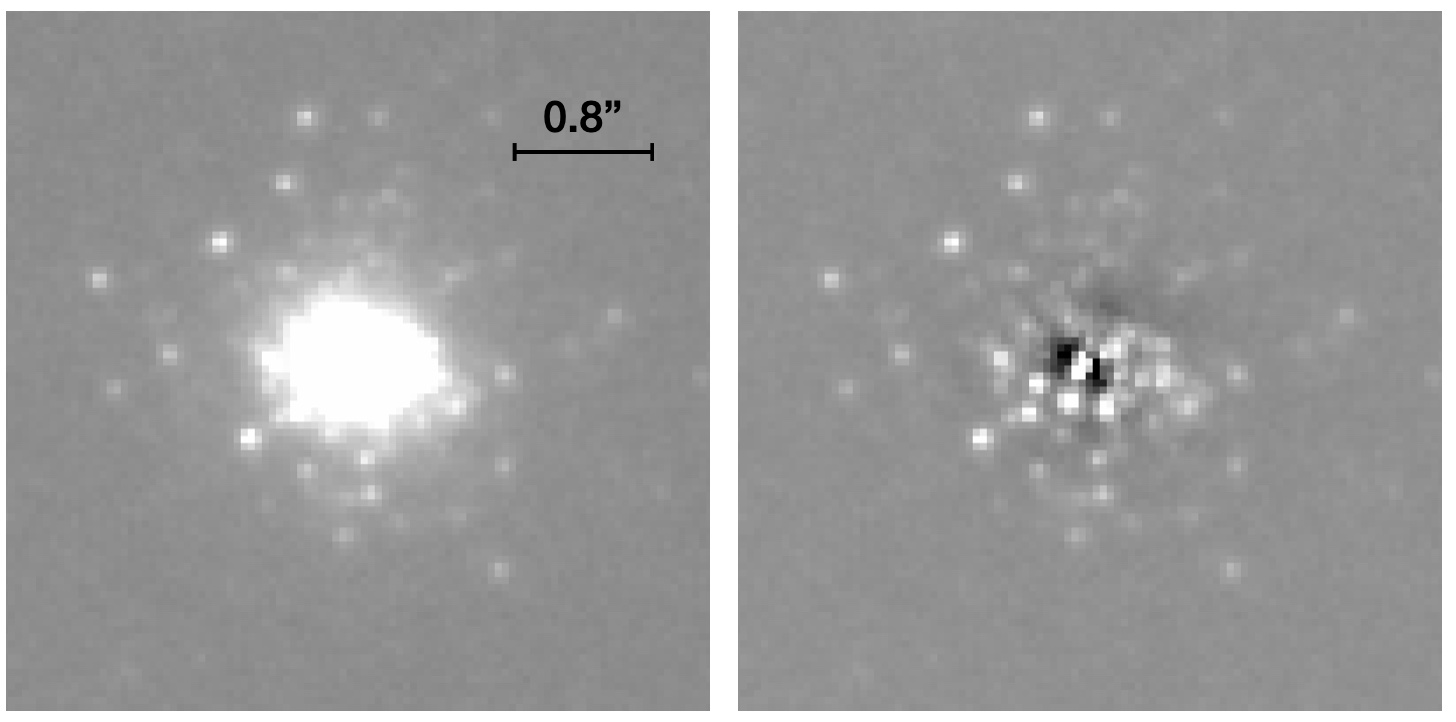}
\caption{Original (left) and residual (right) images with the best-fitting King model subtracted. The large number of resolved giants, including near the center of the cluster, are evident.} 
\label{Fig:sub_img}
\end{figure}

When we fit for the cluster effective radius, we varied the value of the King model concentration parameter (defined as the ratio of the tidal radius to the core radius) using a finely spaced grid that ranged from $c = 10-300$ in steps of one. The best fitting model has a concentration $c = 53$ and an effective radius $r_h = 3.06$ pc. As a King profile is completely specified by two parameters, such a fit implies  a best-fitting projected core radius of $r_c = 0.81$ pc  (Figure \ref{Fig:sub_img}). All the models prefer slightly elliptical fits with a semi-minor to semi-major axis ratio of about 0.92. The best-fitting model has a reduced $\chi^2 = 2.8$, partially due to the presence of many resolved red giants and perhaps partially due to a minor background mismatch in the outer regions of the cluster, relevant given the large fitting radius of 50 pixels = 34.6 pc. Somewhat lower concentrations are possible, the model at $c=25$ having a $\Delta \chi^2 \sim 1$, with a corresponding  $r_h = 2.84$ pc and $r_c = 1.08$ pc . 

At the other end of the distribution, even very large concentrations of $c=300$ or larger give formally reasonable fits with $\Delta \chi^2 < 1$, but the best-fitting structural parameters for these models have large radii. For example, for $c=300$, we find $r_h = 5.30$ pc which would imply $r_c = 0.62$ pc and a tidal radius of 185 pc. Essentially, these fits place more light at large radii where it is poorly constrained by our observations.
Solely considering these data, such models cannot be ruled out, but we also note that a similar combination of high concentration and large radius is essentially absent in the Galactic globular cluster system \citep{1994AJ....108.1292D}, while clusters with parameters close to our best-fit values are common.

Formally, we find an implied $r_c = 0.81^{+0.27}_{-0.19}$ pc. An $r_h < 2.8$ pc is ruled out even in low-concentration models, but as discussed above, models with high concentrations yield formally reasonable fits and have correspondingly larger values of $r_h$, in the range 5--6 pc.  If the globular clusters around M81 are similar to those in the Galaxy, these larger sizes are disfavored. 

As many extragalactic studies assume a fixed concentration index of $c=30$ (e.g., NH11), we also report parameters for this assumed value: $r_h = 2.91$ pc and the inferred  $r_c=1.01$ pc. Further comparisons in this paper to M81 globular clusters or Milky Way globular clusters also use this fixed $c=30$ radius measurement. 

\subsubsection{FRB offset from cluster center}

We used the new HST data to revisit the inferred offset of the FRB from the center of the host cluster \citep{2022Natur.602..585K}. We used 12 stars present in the combined F606W image with measured positions and proper motions from Gaia DR3, advanced to the mean epoch of the HST data, to correct the absolute astrometry of the image. Because this is a relatively small number of stars, the uncertainty in this transformation is $\sim 6$ mas per coordinate. In this frame, the best-fit center determined by {\tt ishape} is
09:57:54.71341, +68:49:00.7818. While this is likely more accurate than the previously published ground-based astrometric positions of the cluster, it is still inferior to the precision of the Gaia DR3 position of the cluster itself, which is listed as 1.6--1.7 mas per coordinate. Our new HST position is offset from the Gaia DR3 position by only 5.8 mas, but given the available information, the Gaia position is still the preferred one to use. The Gaia position implies a projected separation of the cluster center from the FRB of $110\pm2$ mas ($1.92\pm0.03$ pc). For our best-fit model this is within the half-light radius ($\sim 0.63 r_h$). 

\subsection{Optical Photometric Measurements from the HST Images}
 We also measured the integrated magnitude of the candidate in the WFC3 F438W, F606W, and F814W images obtained by the HST 16664 program following the photometric guidelines and calibrations recommended by the WFC3 data handbook \citep{2021wfcd.book....5S}. Formally the uncertainty in the photometry is as low as 0.001 mag in some filters. However, the FRB GC is resolved in the WFC3 images, with individual luminous stars associated with the cluster clearly distinguishable at some wavelengths. We measure the photometry at various radii out to 50 pixels in order to optimize the S/N vs the flux, and estimate that the systematic uncertainty in the photometry is at least 0.02 mag. Using Vegamag zeropoints in order to be consistent with the ground based BVR observations we measure F438W=20.75, F606W=19.59, F814W=18.73 $\pm$ 0.02 mag, values generally consistent with those from previous works. The photometry of the FRB GC places the cluster securely within the expected magnitude and color range for a globular cluster  
\citep[e.g.,][]{Rhode2001}. We do not expect that a stellar population analysis using this photometry would produce a meaningfully different inferred stellar mass or metallicity than previous work (e.g., \citealt{2022Natur.602..585K}), though it could be illuminating to construct a color-magnitude diagram from the HST data in the future, an effort which is outside the scope of the present paper.

\begin{figure*}
\includegraphics[width=0.95\textwidth]{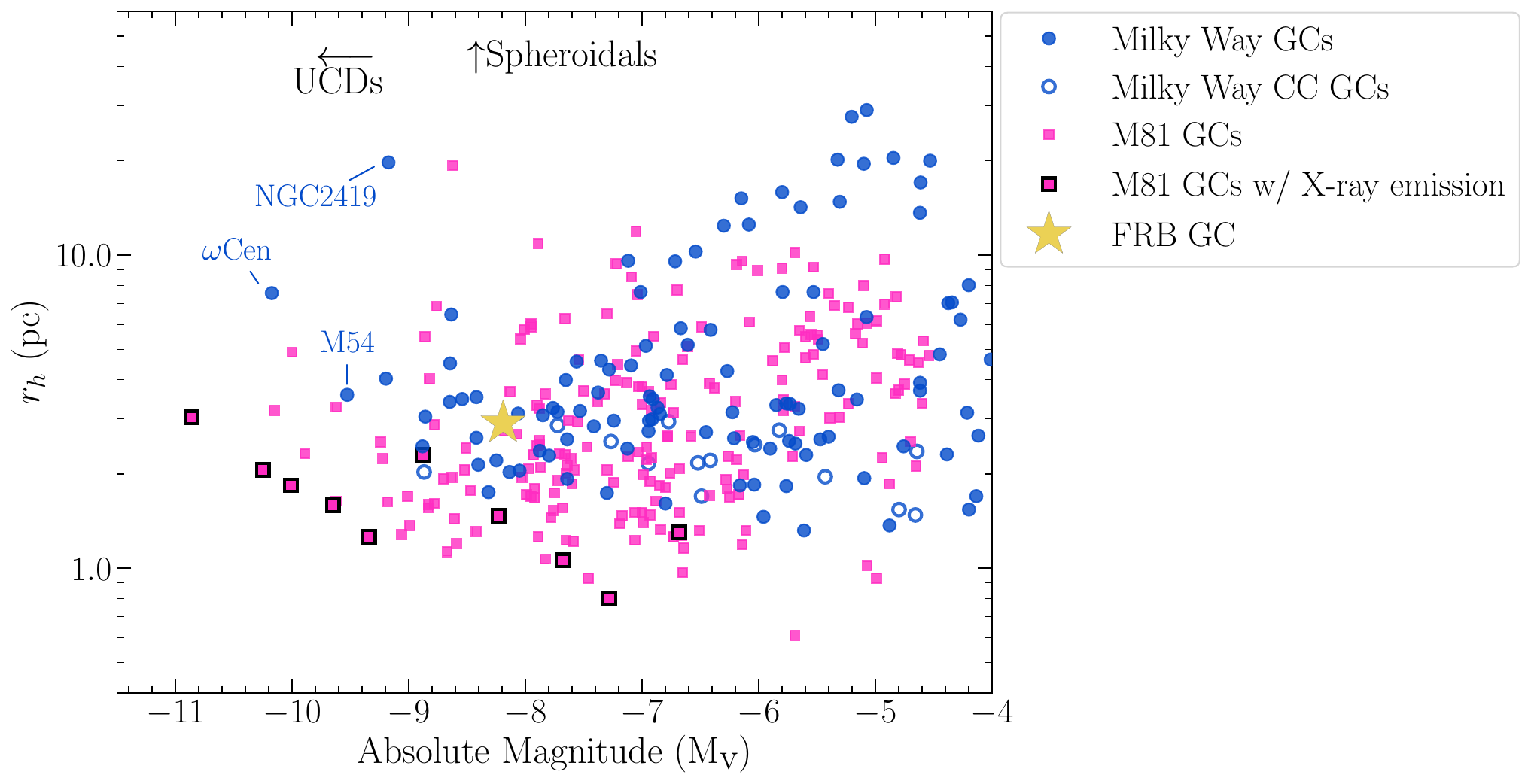}
\caption{Scatter plot of half-light radii versus absolute magnitude for globular clusters in the Milky Way and M81, in the context of star clusters and dwarf galaxies \citep[e.g., see][]{Tolstoy2009}. Core-collapsed globular clusters in the Milky Way are denoted by hollow circles,  and globular clusters in M81 with bright X-ray sources are denoted with black margins. The FRB GC in M81 does not appear to be exceptional in its properties compared to these other cluster populations.} 
\label{Fig:rhVmag_plot}
\end{figure*}

\section{Comparisons Between M81 and Milky Way globular clusters} \label{sec:comparison}

In this section we compare a few dynamical and structural parameters measured for the M81 FRB host GC with those measured for Galactic clusters and globular cluster candidates in M81 (with measurements from NH11; 85 classified as ``confirmed'' and 125 classified as ``good'' candidates by NH11). We note that \cite{Pan2022} et al. suggest that the HST selected candidates of NH11 have an estimated contamination rate of 8\% for the bright clusters.

\subsection{Structural Properties}
We compared the FRB GC optical absolute V magnitude ($-8.19$, this work), and half-light radius (2.91 parsec for a King concentration of 30, this work) to a sample of 160 Galactic globular clusters from \cite{2020PASA...37...46B}, and 210 M81 globular clusters from NH11. As shown in Figure \ref{Fig:rhVmag_plot}, the FRB GC does not appear to be structurally exceptional.

\subsection{Relative Stellar Encounter Rate}

To estimate encounter rates for extragalactic globular clusters in M81, we use the proxy from equation (5) of \cite{Sivakoff07}: 
\begin{equation}
    \Gamma_{h} \equiv \left(\frac{M}{2\pi \Msun}\right)^{\frac{3}{2}}\left(\frac{r_h}{1pc}\right)^{\frac{-5}{2}},
\end{equation}

where M is the optical mass of the cluster, and r is the observed half-light radius.

Based on our updated measurements, and using the distance of 3.6 Mpc, the absolute V magnitude of the FRB GC is -8.19. We convert optical magnitude to cluster mass by adopting the median V magnitude globular cluster mass-to-light-ratio of Galactic globular clusters \citep[1.83;][]{2020PASA...37...46B}. Assuming a V magnitude of 4.81 for the Sun, we find the cluster mass of the FRB GC to be 2.9 $\times 10^{5}$ \Msun. We perform the same conversion to the V magnitudes of the NH11 clusters. 

\begin{figure}
    \centering
    \includegraphics[width=0.48\textwidth]{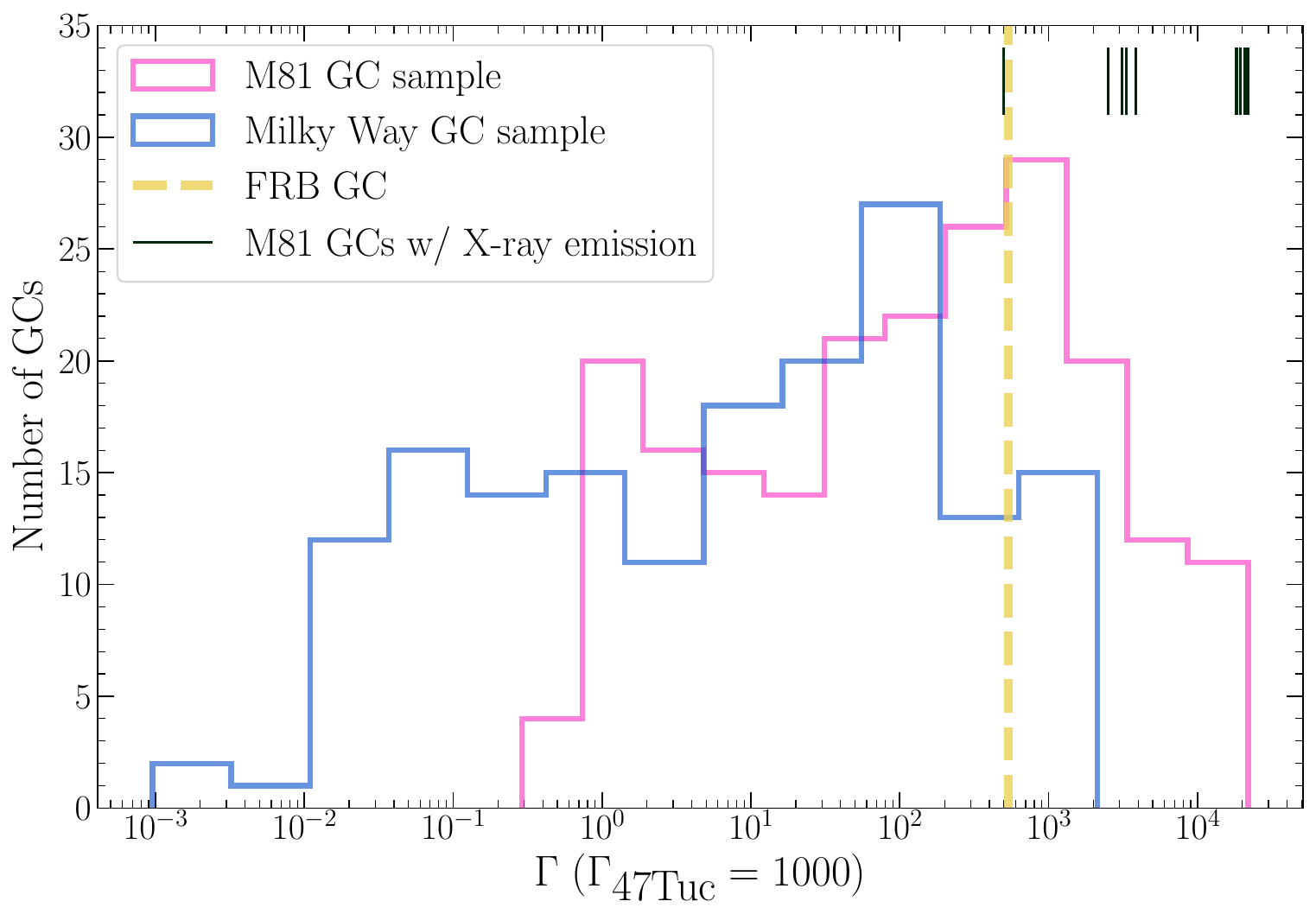}
    \caption{Relative encounter rates for globular clusters as they would be observed in the M81 system, normalized assuming 47Tuc has an encounter rate of 1000.}
    \label{fig:relER}
\end{figure}

We perform the same conversion to optical mass and encounter rate calculation using the \cite{2020PASA...37...46B} V magnitudes and half-light radii of Galactic globular clusters. We caution that this exercise is meant to demonstrate the encounter rates of Galactic globular clusters if they were observed in the M81 system. For ease of comparison, we normalize all estimated encounter rates to that of 47 Tuc, assuming $\Gamma_\textrm{47Tuc}=1000$. As demonstrated in Figure \ref{fig:relER}, the FRB GC has an encounter rate that is $\sim 50\%$ of that of 47 Tuc. This indicates that in principle, a dynamical formation may be a plausible formation channel for the progenitor of FRB20200120E. However, we note that the comparison data that we are using - i.e. the   measurements of Galactic globular clusters from a range of studies and data from extragalactic globular clusters that are derived from HST observations --  have very different observational limits, selection biases, uncertainties, and contamination rates. Coupled with the differences in how cluster radii are estimated in these different data sets, comparisons of the absolute interaction rates of M81 and Milky Way GCs are subject to systematic uncertainties and should be considered indicative at best.

Considering only the M81 GC candidates, we note that the effective radius of the FRB host is indistinguishable from the median of the sample, but the luminosity (mass) is $\approx$0.7 mag brighter than the peak of the globular cluster luminosity function (Fig 3). It is this latter difference that manifests itself in $\Gamma$ and suggests that the FRB resides in a GC that has a moderately higher interaction rate than the median (Fig 4). This is the consequence of an intriguing characteristic of GCs; the lack of a mass-radius relationship. This results in the stellar density and dynamical interaction rate in globular clusters to be strongly correlated to the mass of a cluster. A corollary to this feature is that the vast majority of dynamical interactions in globular cluster systems occur in the most massive GCs. Therefore, the offset of the FRB GC from the median interaction rate hints at the possibility of the FRB being dynamically formed, but it is not conclusive.

We note that \citealt{Kremer21, Kremerwd} predict that white dwarf mergers should occur mainly in clusters undergoing core collapse or in the post-core collapse phase. While our analysis does not allow us to determine whether the M81 FRB GC has reached core collapse, it is worth noting that the vast majority of post-core collapse clusters in the Milky Way are located at small galactocentric distances  \citep[see e.g.,][]{1989ApJ...339..904C, 1994AJ....108.1292D} and they are all closer to the Galactic  center than the distance of the M81 FRB GC from the center of M81. Note that for the M81 FRB GC, 20.5 kpc is the projected galactocentric distance so the actual 3D distance is likely to be larger.

 We also  note  that the FRB GC is located at a much larger galactocentric distance than the rest of the M81 sample and studies of the globular cluster systems of the Milky Way and other Galaxies suggest a radial trend in GC sizes. A high resolution survey of other clusters at the galactocentric distance of the FRB GC will help place the relative $\Gamma$ of this GC  and the importance of dynamical effects in better context.

\subsection{Comparison to X-ray Binary Hosting Globular Clusters in M81}
While the FRB GC itself has been observed to show no evidence of X-ray emission \citep{2023arXiv230810930P}, it is nevertheless worth comparing the cluster properties to those of other GCs in M81 which show X-ray emission, as bright X-ray activity from X-ray binaries can be transient. Several globular clusters in M81 are known to host X-ray binaries \citep{2023ApJ...947...31H}. We cross-matched the NH11 globular clusters with the Chandra Source Catalog Version 2.1 \citep{2020AAS...23515405E}, and found that ten of the NH11 globular clusters have significant (at $\geq3\sigma$) X-ray counterparts detected by \textit{Chandra}, with X-ray luminosities spanning from 2.0 $\times$ 10$^{37}$ erg/s to 5.2 $\times 10^{38}$ erg/s. We note that the majority (9) of the X-ray hosting clusters are at projected distances within roughly 3 kpc of the galaxy center, and one is 11.1 kpc away. 

The mean V magnitude for the entire NH11 sample is $-7.0$ mag, with a standard deviation of 1.4 mag. The mean half-light radius is 3.5 pc with a standard deviation of 2.5 pc. For the sample of NH11 globular cluster candidates with X-ray counterparts, the mean V magnitude is $-9.0$ mag, with a 1.1 mag standard deviation. The mean half-light radius of the X-ray-detected NH11 globular clusters is 1.7 pc, with a standard deviation of 0.6 pc. While there is not a clear trend in absolute magnitude, beyond the FRB GC being on the faint end of the M81 X-ray binary hosting sample, it is clear that the X-ray detected globular clusters are on average more compact than the FRB GC.

\section{Summary and Discussion} \label{sec:summary}
The M81 globular cluster that hosts FRB 20200120E offers a rare opportunity to study the environment that produced such an extreme emission source. Our analysis of the cluster structural parameters in HST F606W contrains the effective radius to $r_h = 3.06$ pc and a moderate King model concentration of $c=53$. This implies 
a core radius of 0.81 pc. Our new photometric measurements of the cluster are F438W=20.75, F606W=19.59, and F814W=18.73 $\pm$ 0.02 mag. 

We compared the optical properties of the host cluster to other clusters in M81, and to Galactic globular clusters. We find that the FRB GC is not observed to be unique or extreme in comparison to other globular clusters in M81, or to Galactic globular clusters.  The FRB resides in a globular cluster host that has a moderately higher than median stellar encounter rate. This implies that dynamical interactions are a plausible formation path for the FRB progenitor, but given the uncertainties in the available measurements and comparison data we cannot confirm this formation channel. 

\vspace{5mm}
\facilities{\textit{HST}, \textit{Chandra}}

\software{astropy \citep{Robitaille13},  baolab \citep{Larsen99,2014ascl.soft03013L}, DrizzlePAC \citep{2021AAS...23821602H}, IRAF \citep{1986SPIE..627..733T,1993ASPC...52..173T}, Matplotlib \citep{Hunter07}, NumPy \citep{harris20}, Pandas \citep{Mckinney10}}

\begin{acknowledgments}
We thank the referee for helpful comments which improved the manuscript. The authors thank S{\o}ren Larsen and Andy Fruchter for helpful discussions, and the referee for useful suggestions that helped improve this paper. KCD acknowledges support for this work was provided by NASA through the NASA Hubble Fellowship grant
HST-HF2-51528 awarded by the Space Telescope Science Institute, which is operated by the Association of Universities for Research in Astronomy, Inc., for NASA, under contract NAS5–26555. JS acknowledges support from 
NASA grant 80NSSC21K0628 and the Packard Foundation. 
We acknowledge extensive use of NASA's Astrophysics Data System Bibliographic Services, Arxiv, and SIMBAD \citep{Wenger2000}.
\end{acknowledgments}

\bibliography{draft_arxiv}{}

\begin{thebibliography}{}
\expandafter\ifx\csname natexlab\endcsname\relax\def\natexlab#1{#1}\fi
\providecommand{\url}[1]{\href{#1}{#1}}
\providecommand{\dodoi}[1]{doi:~\href{http://doi.org/#1}{\nolinkurl{#1}}}
\providecommand{\doeprint}[1]{\href{http://ascl.net/#1}{\nolinkurl{http://ascl.net/#1}}}
\providecommand{\doarXiv}[1]{\href{https://arxiv.org/abs/#1}{\nolinkurl{https://arxiv.org/abs/#1}}}

\bibitem[{{Astropy Collaboration} {et~al.}(2013){Astropy Collaboration},
  {Robitaille}, {Tollerud}, {Greenfield}, {Droettboom}, {Bray}, {Aldcroft},
  {Davis}, {Ginsburg}, {Price-Whelan}, {Kerzendorf}, {Conley}, {Crighton},
  {Barbary}, {Muna}, {Ferguson}, {Grollier}, {Parikh}, {Nair}, {Unther},
  {Deil}, {Woillez}, {Conseil}, {Kramer}, {Turner}, {Singer}, {Fox}, {Weaver},
  {Zabalza}, {Edwards}, {Azalee Bostroem}, {Burke}, {Casey}, {Crawford},
  {Dencheva}, {Ely}, {Jenness}, {Labrie}, {Lim}, {Pierfederici}, {Pontzen},
  {Ptak}, {Refsdal}, {Servillat}, \& {Streicher}}]{Robitaille13}
{Astropy Collaboration}, {Robitaille}, T.~P., {Tollerud}, E.~J., {et~al.} 2013,
  \aap, 558, A33, \dodoi{10.1051/0004-6361/201322068}

\bibitem[{{Bahramian} {et~al.}(2013){Bahramian}, {Heinke}, {Sivakoff}, \&
  {Gladstone}}]{Bahramian13}
{Bahramian}, A., {Heinke}, C.~O., {Sivakoff}, G.~R., \& {Gladstone}, J.~C.
  2013, \apj, 766, 136, \dodoi{10.1088/0004-637X/766/2/136}

\bibitem[{{Bannister} {et~al.}(2019){Bannister}, {Deller}, {Phillips},
  {Macquart}, {Prochaska}, {Tejos}, {Ryder}, {Sadler}, {Shannon}, {Simha},
  {Day}, {McQuinn}, {North-Hickey}, {Bhandari}, {Arcus}, {Bennert}, {Burchett},
  {Bouwhuis}, {Dodson}, {Ekers}, {Farah}, {Flynn}, {James}, {Kerr}, {Lenc},
  {Mahony}, {O{\textquoteright}Meara}, {Os{\l}owski}, {Qiu}, {Treu}, {U},
  {Bateman}, {Bock}, {Bolton}, {Brown}, {Bunton}, {Chippendale}, {Cooray},
  {Cornwell}, {Gupta}, {Hayman}, {Kesteven}, {Koribalski}, {MacLeod},
  {McClure-Griffiths}, {Neuhold}, {Norris}, {Pilawa}, {Qiao}, {Reynolds},
  {Roxby}, {Shimwell}, {Voronkov}, \& {Wilson}}]{Bannister2019}
{Bannister}, K.~W., {Deller}, A.~T., {Phillips}, C., {et~al.} 2019, Science,
  365, 565, \dodoi{10.1126/science.aaw5903}

\bibitem[{{Baumgardt} {et~al.}(2020){Baumgardt}, {Sollima}, \&
  {Hilker}}]{2020PASA...37...46B}
{Baumgardt}, H., {Sollima}, A., \& {Hilker}, M. 2020, \pasa, 37, e046,
  \dodoi{10.1017/pasa.2020.38}

\bibitem[{{Bhandari} {et~al.}(2022){Bhandari}, {Heintz}, {Aggarwal}, {Marnoch},
  {Day}, {Sydnor}, {Burke-Spolaor}, {Law}, {Xavier Prochaska}, {Tejos},
  {Bannister}, {Butler}, {Deller}, {Ekers}, {Flynn}, {Fong}, {James}, {Lazio},
  {Luo}, {Mahony}, {Ryder}, {Sadler}, {Shannon}, {Han}, {Lee}, \&
  {Zhang}}]{Bhandari+22}
{Bhandari}, S., {Heintz}, K.~E., {Aggarwal}, K., {et~al.} 2022, \aj, 163, 69,
  \dodoi{10.3847/1538-3881/ac3aec}

\bibitem[{{Chatterjee} {et~al.}(2017){Chatterjee}, {Law}, {Wharton},
  {Burke-Spolaor}, {Hessels}, {Bower}, {Cordes}, {Tendulkar}, {Bassa},
  {Demorest}, {Butler}, {Seymour}, {Scholz}, {Abruzzo}, {Bogdanov}, {Kaspi},
  {Keimpema}, {Lazio}, {Marcote}, {McLaughlin}, {Paragi}, {Ransom}, {Rupen},
  {Spitler}, \& {van Langevelde}}]{VLAlocalisation}
{Chatterjee}, S., {Law}, C.~J., {Wharton}, R.~S., {et~al.} 2017, \nat, 541, 58,
  \dodoi{10.1038/nature20797}

\bibitem[{{Chernoff} \& {Djorgovski}(1989)}]{1989ApJ...339..904C}
{Chernoff}, D.~F., \& {Djorgovski}, S. 1989, \apj, 339, 904,
  \dodoi{10.1086/167344}

\bibitem[{{Chies-Santos} {et~al.}(2022){Chies-Santos}, {de Souza}, {Caso},
  {Ennis}, {de Souza}, {Barbosa}, {Chen}, {Javier Cenarro}, {Ederoclite},
  {Crist{\'o}bal-Hornillos}, {Hern{\'a}ndez-Monteagudo}, {L{\'o}pez-Sanjuan},
  {Mar{\'\i}n-Franch}, {Moles}, {Varela}, {V{\'a}zquez Rami{\'o}}, {Dupke},
  {Sodr{\'e}}, \& {Angulo}}]{2022MNRAS.516.1320C}
{Chies-Santos}, A.~L., {de Souza}, R.~S., {Caso}, J.~P., {et~al.} 2022, \mnras,
  516, 1320, \dodoi{10.1093/mnras/stac2002}

\bibitem[{{Dage} {et~al.}(2020){Dage}, {Zepf}, {Thygesen}, {Bahramian},
  {Kundu}, {Maccarone}, {Peacock}, \& {Strader}}]{2020MNRAS.497..596D}
{Dage}, K.~C., {Zepf}, S.~E., {Thygesen}, E., {et~al.} 2020, \mnras, 497, 596,
  \dodoi{10.1093/mnras/staa1963}

\bibitem[{{Djorgovski} \& {Meylan}(1994)}]{1994AJ....108.1292D}
{Djorgovski}, S., \& {Meylan}, G. 1994, \aj, 108, 1292, \dodoi{10.1086/117154}

\bibitem[{{Evans} {et~al.}(2020){Evans}, {Primini}, {Miller}, {Evans}, {Allen},
  {Anderson}, {Becker}, {Budynkiewicz}, {Burke}, {Chen}, {Civano}, {D'Abrusco},
  {Doe}, {Fabbiano}, {Martinez Galarza}, {Gibbs}, {Glotfelty}, {Graessle},
  {Grier}, {Hain}, {Hall}, {Harbo}, {Houck}, {Lauer}, {Laurino}, {Lee},
  {McCollough}, {McDowell}, {McLaughlin}, {Morgan}, {Mossman}, {Nguyen},
  {Nichols}, {Nowak}, {Paxson}, {Perdikeas}, {Plummer}, {Rots},
  {Siemiginowska}, {Sundheim}, {Thong}, {Tibbetts}, {Van Stone}, {Winkelman},
  \& {Zografou}}]{2020AAS...23515405E}
{Evans}, I.~N., {Primini}, F.~A., {Miller}, J.~B., {et~al.} 2020, in American
  Astronomical Society Meeting Abstracts, Vol. 235, American Astronomical
  Society Meeting Abstracts \#235, 154.05

\bibitem[{{Falcke} \& {Rezzolla}(2014)}]{FalckeRezollaBlitzar}
{Falcke}, H., \& {Rezzolla}, L. 2014, \aap, 562, A137,
  \dodoi{10.1051/0004-6361/201321996}

\bibitem[{{Gordon} {et~al.}(2023){Gordon}, {Fong}, {Kilpatrick}, {Eftekhari},
  {Leja}, {Prochaska}, {Nugent}, {Bhandari}, {Blanchard}, {Caleb}, {Day},
  {Deller}, {Dong}, {Glowacki}, {Gourdji}, {Mannings}, {Mahoney}, {Marnoch},
  {Miller}, {Paterson}, {Rastinejad}, {Ryder}, {Sadler}, {Scott}, {Sears},
  {Shannon}, {Simha}, {Stappers}, \& {Tejos}}]{Gordon2023}
{Gordon}, A.~C., {Fong}, W.-f., {Kilpatrick}, C.~D., {et~al.} 2023, arXiv
  e-prints, arXiv:2302.05465, \dodoi{10.48550/arXiv.2302.05465}

\bibitem[{Harris {et~al.}(2020)Harris, Millman, van~der Walt, Gommers,
  Virtanen, Cournapeau, Wieser, Taylor, Berg, Smith, Kern, Picus, Hoyer, van
  Kerkwijk, Brett, Haldane, del R{'{\i}}o, Wiebe, Peterson,
  G{'{e}}rard-Marchant, Sheppard, Reddy, Weckesser, Abbasi, Gohlke, \&
  Oliphant}]{harris20}
Harris, C.~R., Millman, K.~J., van~der Walt, S.~J., {et~al.} 2020, Nature, 585,
  357, \dodoi{10.1038/s41586-020-2649-2}

\bibitem[{{Heinke} {et~al.}(2003){Heinke}, {Grindlay}, {Lugger}, {Cohn},
  {Edmonds}, {Lloyd}, \& {Cool}}]{Heinke03d}
{Heinke}, C.~O., {Grindlay}, J.~E., {Lugger}, P.~M., {et~al.} 2003, \apj, 598,
  501

\bibitem[{{Hills}(1976)}]{Hills76}
{Hills}, J.~G. 1976, \mnras, 175, 1P

\bibitem[{{Hoffmann} {et~al.}(2021){Hoffmann}, {Mack}, {Avila}, {Martlin},
  {Cohen}, \& {Bajaj}}]{2021AAS...23821602H}
{Hoffmann}, S.~L., {Mack}, J., {Avila}, R., {et~al.} 2021, in American
  Astronomical Society Meeting Abstracts, Vol.~53, American Astronomical
  Society Meeting Abstracts, 216.02

\bibitem[{{Hunt} {et~al.}(2023){Hunt}, {Gallo}, {Chandar}, {Mok}, \&
  {Prestwich}}]{2023ApJ...947...31H}
{Hunt}, Q., {Gallo}, E., {Chandar}, R., {Mok}, A., \& {Prestwich}, A. 2023,
  \apj, 947, 31, \dodoi{10.3847/1538-4357/ac78da}

\bibitem[{Hunter(2007)}]{Hunter07}
Hunter, J.~D. 2007, Computing In Science \& Engineering, 9, 90,
  \dodoi{10.1109/MCSE.2007.55}

\bibitem[{{James} {et~al.}(2020){James}, {Os{\l}owski}, {Flynn}, {Kumar},
  {Bannister}, {Bhandari}, {Farah}, {Kerr}, {Lorimer}, {Macquart}, {Ng},
  {Phillips}, {Price}, {Qiu}, {Shannon}, \& {Spiewak}}]{James2020a_followup}
{James}, C.~W., {Os{\l}owski}, S., {Flynn}, C., {et~al.} 2020, \mnras, 495,
  2416, \dodoi{10.1093/mnras/staa1361}

\bibitem[{{Jord{\'a}n} {et~al.}(2005){Jord{\'a}n}, {C{\^o}t{\'e}}, {Blakeslee},
  {Ferrarese}, {McLaughlin}, {Mei}, {Peng}, {Tonry}, {Merritt},
  {Milosavljevi{\'c}}, {Sarazin}, {Sivakoff}, \& {West}}]{2005ApJ...634.1002J}
{Jord{\'a}n}, A., {C{\^o}t{\'e}}, P., {Blakeslee}, J.~P., {et~al.} 2005, \apj,
  634, 1002, \dodoi{10.1086/497092}

\bibitem[{{King}(1962)}]{King62}
{King}, I. 1962, \aj, 67, 471, \dodoi{10.1086/108756}

\bibitem[{{King}(1966)}]{King66}
{King}, I.~R. 1966, \aj, 71, 64

\bibitem[{{Kirsten} {et~al.}(2022){Kirsten}, {Marcote}, {Nimmo}, {Hessels},
  {Bhardwaj}, {Tendulkar}, {Keimpema}, {Yang}, {Snelders}, {Scholz},
  {Pearlman}, {Law}, {Peters}, {Giroletti}, {Paragi}, {Bassa}, {Hewitt},
  {Bach}, {Bezrukovs}, {Burgay}, {Buttaccio}, {Conway}, {Corongiu}, {Feiler},
  {Forss{\'e}n}, {Gawro{\'n}ski}, {Karuppusamy}, {Kharinov}, {Lindqvist},
  {Maccaferri}, {Melnikov}, {Ould-Boukattine}, {Possenti}, {Surcis}, {Wang},
  {Yuan}, {Aggarwal}, {Anna-Thomas}, {Bower}, {Blaauw}, {Burke-Spolaor},
  {Cassanelli}, {Clarke}, {Fonseca}, {Gaensler}, {Gopinath}, {Kaspi}, {Kassim},
  {Lazio}, {Leung}, {Li}, {Lin}, {Masui}, {Mckinven}, {Michilli}, {Mikhailov},
  {Ng}, {Orbidans}, {Pen}, {Petroff}, {Rahman}, {Ransom}, {Shin}, {Smith},
  {Stairs}, \& {Vlemmings}}]{2022Natur.602..585K}
{Kirsten}, F., {Marcote}, B., {Nimmo}, K., {et~al.} 2022, \nat, 602, 585,
  \dodoi{10.1038/s41586-021-04354-w}

\bibitem[{{Kremer} {et~al.}(2023{\natexlab{a}}){Kremer}, {Fuller}, {Piro}, \&
  {Ransom}}]{Kremerwd}
{Kremer}, K., {Fuller}, J., {Piro}, A.~L., \& {Ransom}, S.~M.
  2023{\natexlab{a}}, \mnras, 525, L22, \dodoi{10.1093/mnrasl/slad088}

\bibitem[{{Kremer} {et~al.}(2023{\natexlab{b}}){Kremer}, {Li}, {Lu}, {Piro}, \&
  {Zhang}}]{Kremerm87}
{Kremer}, K., {Li}, D., {Lu}, W., {Piro}, A.~L., \& {Zhang}, B.
  2023{\natexlab{b}}, \apj, 944, 6, \dodoi{10.3847/1538-4357/acabbf}

\bibitem[{{Kremer} {et~al.}(2021){Kremer}, {Piro}, \& {Li}}]{Kremer21}
{Kremer}, K., {Piro}, A.~L., \& {Li}, D. 2021, \apjl, 917, L11,
  \dodoi{10.3847/2041-8213/ac13a0}

\bibitem[{{Kundu} {et~al.}(1999){Kundu}, {Whitmore}, {Sparks}, {Macchetto},
  {Zepf}, \& {Ashman}}]{1999ApJ...513..733K}
{Kundu}, A., {Whitmore}, B.~C., {Sparks}, W.~B., {et~al.} 1999, \apj, 513, 733,
  \dodoi{10.1086/306865}

\bibitem[{{Larsen}(1999)}]{Larsen99}
{Larsen}, S.~S. 1999, \aaps, 139, 393, \dodoi{10.1051/aas:1999509}

\bibitem[{{Larsen}(2014)}]{2014ascl.soft03013L}
---. 2014, {BAOlab: Image processing program}, Astrophysics Source Code
  Library, record ascl:1403.013.
\newblock \doeprint{1403.013}

\bibitem[{{Lee-Waddell} {et~al.}(2023){Lee-Waddell}, {James}, {Ryder},
  {Mahony}, {Bahramian}, {Koribalski}, {Kumar}, {Marnoch}, {North-Hickey},
  {Sadler}, {Shannon}, {Tejos}, {Thorne}, {Wang}, \&
  {Wayth}}]{Lee-Waddell_2023}
{Lee-Waddell}, K., {James}, C.~W., {Ryder}, S.~D., {et~al.} 2023, arXiv
  e-prints, arXiv:2305.17960, \dodoi{10.48550/arXiv.2305.17960}

\bibitem[{{Lin} {et~al.}(2023){Lin}, {Scholz}, {Ng}, {Pen}, {Bhardwaj},
  {Chawla}, {Curtin}, {Sand}, {Tendulkar}, {Andersen}, {Bandura}, {Cassanelli},
  {Cook}, {Dobbs}, {Dong}, {Eadie}, {Fonseca}, {Gaensler}, {Giri},
  {Herrera-Martin}, {Kaczmarek}, {Kania}, {Kaspi}, {Khairy}, {Lanman}, {Leung},
  {Li}, {Masui}, {Mena-Parra}, {Meyers}, {Michilli}, {Milutinovic}, {Pearlman},
  {Pleunis}, {Rafiei-Ravandi}, {Rahman}, {Sanghavi}, {Shin}, {Smith}, {Stairs},
  {Stenning}, {Vanderlinde}, \& {Wulf}}]{2023CHIEMFarSidelobes}
{Lin}, H.-H., {Scholz}, P., {Ng}, C., {et~al.} 2023, arXiv e-prints,
  arXiv:2307.05261, \dodoi{10.48550/arXiv.2307.05261}

\bibitem[{{Lorimer} {et~al.}(2007){Lorimer}, {Bailes}, {McLaughlin},
  {Narkevic}, \& {Crawford}}]{Lorimer2007}
{Lorimer}, D.~R., {Bailes}, M., {McLaughlin}, M.~A., {Narkevic}, D.~J., \&
  {Crawford}, F. 2007, Science, 318, 777, \dodoi{10.1126/science.1147532}

\bibitem[{{Ma} {et~al.}(2007){Ma}, {Burstein}, {Fan}, {Zhou}, {Chen}, {Jiang},
  {Wu}, \& {Wu}}]{2007PASP..119.1085M}
{Ma}, J., {Burstein}, D., {Fan}, Z., {et~al.} 2007, \pasp, 119, 1085,
  \dodoi{10.1086/522870}

\bibitem[{{Marcote} {et~al.}(2020){Marcote}, {Nimmo}, {Hessels}, {Tendulkar},
  {Bassa}, {Paragi}, {Keimpema}, {Bhardwaj}, {Karuppusamy}, {Kaspi}, {Law},
  {Michilli}, {Aggarwal}, {Andersen}, {Archibald}, {Bandura}, {Bower}, {Boyle},
  {Brar}, {Burke-Spolaor}, {Butler}, {Cassanelli}, {Chawla}, {Demorest},
  {Dobbs}, {Fonseca}, {Giri}, {Good}, {Gourdji}, {Josephy}, {Kirichenko},
  {Kirsten}, {Landecker}, {Lang}, {Lazio}, {Li}, {Lin}, {Linford}, {Masui},
  {Mena-Parra}, {Naidu}, {Ng}, {Patel}, {Pen}, {Pleunis}, {Rafiei-Ravandi},
  {Rahman}, {Renard}, {Scholz}, {Siegel}, {Smith}, {Stairs}, {Vanderlinde}, \&
  {Zwaniga}}]{MarcoteRepeaterLocalisation2020}
{Marcote}, B., {Nimmo}, K., {Hessels}, J.~W.~T., {et~al.} 2020, \nat, 577, 190,
  \dodoi{10.1038/s41586-019-1866-z}

\bibitem[{McKinney(2010)}]{Mckinney10}
McKinney, W. 2010, in Proceedings of the 9th Python in Science Conference, ed.
  S.~van~der Walt \& J.~Millman, 51 -- 56

\bibitem[{{Metzger} {et~al.}(2017){Metzger}, {Berger}, \&
  {Margalit}}]{Metzger2017magnetar}
{Metzger}, B.~D., {Berger}, E., \& {Margalit}, B. 2017, \apj, 841, 14,
  \dodoi{10.3847/1538-4357/aa633d}

\bibitem[{{Michilli} {et~al.}(2018){Michilli}, {Seymour}, {Hessels}, {Spitler},
  {Gajjar}, {Archibald}, {Bower}, {Chatterjee}, {Cordes}, {Gourdji}, {Heald},
  {Kaspi}, {Law}, {Sobey}, {Adams}, {Bassa}, {Bogdanov}, {Brinkman},
  {Demorest}, {Fernand ez}, {Hellbourg}, {Lazio}, {Lynch}, {Maddox}, {Marcote},
  {McLaughlin}, {Paragi}, {Ransom}, {Scholz}, {Siemion}, {Tendulkar}, {van
  Rooy}, {Wharton}, \& {Whitlow}}]{Michilli2018_121102}
{Michilli}, D., {Seymour}, A., {Hessels}, J.~W.~T., {et~al.} 2018, \nat, 553,
  182, \dodoi{10.1038/nature25149}

\bibitem[{{Nantais} {et~al.}(2011){Nantais}, {Huchra}, {Zezas}, {Gazeas}, \&
  {Strader}}]{2011AJ....142..183N}
{Nantais}, J.~B., {Huchra}, J.~P., {Zezas}, A., {Gazeas}, K., \& {Strader}, J.
  2011, \aj, 142, 183, \dodoi{10.1088/0004-6256/142/6/183}

\bibitem[{{Niu} {et~al.}(2022){Niu}, {Aggarwal}, {Li}, {Zhang}, {Chatterjee},
  {Tsai}, {Yu}, {Law}, {Burke-Spolaor}, {Cordes}, {Zhang}, {Ocker}, {Yao},
  {Wan}, {Feng}, {Niino}, {Bochenek}, {Cruces}, {Connor}, {Jiang}, {Dai},
  {Luo}, {Li}, {Miao}, {Niu}, {Anna-Thomas}, {Sydnor}, {Stern}, {Wang}, {Yuan},
  {Yue}, {Zhou}, {Yan}, {Zhu}, \& {Zhang}}]{Niu2022}
{Niu}, C.~H., {Aggarwal}, K., {Li}, D., {et~al.} 2022, \nat, 606, 873,
  \dodoi{10.1038/s41586-022-04755-5}

\bibitem[{{Pan} {et~al.}(2022){Pan}, {Bell}, {Smercina}, {Price}, {Slater},
  {Bailin}, {de Jong}, {D'Souza}, {Jang}, \& {Monachesi}}]{Pan2022}
{Pan}, J., {Bell}, E.~F., {Smercina}, A., {et~al.} 2022, \mnras, 515, 48,
  \dodoi{10.1093/mnras/stac1638}

\bibitem[{{Peacock} {et~al.}(2012){Peacock}, {Zepf}, {Kundu}, {Maccarone},
  {Rhode}, {Salzer}, {Waters}, {Ciardullo}, {Gronwall}, \&
  {Stern}}]{2012ApJ...759..126P}
{Peacock}, M.~B., {Zepf}, S.~E., {Kundu}, A., {et~al.} 2012, \apj, 759, 126,
  \dodoi{10.1088/0004-637X/759/2/126}

\bibitem[{{Pearlman} {et~al.}(2023){Pearlman}, {Scholz}, {Bethapudi},
  {Hessels}, {Kaspi}, {Kirsten}, {Nimmo}, {Spitler}, {Fonseca}, {Meyers},
  {Stairs}, {Tan}, {Bhardwaj}, {Chatterjee}, {Cook}, {Curtin}, {Dong},
  {Eftekhari}, {Gaensler}, {G{\"u}ver}, {Kaczmarek}, {Leung}, {Masui},
  {Michilli}, {Prince}, {Sand}, {Shin}, {Smith}, \&
  {Tendulkar}}]{2023arXiv230810930P}
{Pearlman}, A.~B., {Scholz}, P., {Bethapudi}, S., {et~al.} 2023, arXiv
  e-prints, arXiv:2308.10930, \dodoi{10.48550/arXiv.2308.10930}

\bibitem[{{Perelmuter} {et~al.}(1995){Perelmuter}, {Brodie}, \&
  {Huchra}}]{1995AJ....110..620P}
{Perelmuter}, J.-M., {Brodie}, J.~P., \& {Huchra}, J.~P. 1995, \aj, 110, 620,
  \dodoi{10.1086/117547}

\bibitem[{{Perelmuter} \& {Racine}(1995)}]{1995AJ....109.1055P}
{Perelmuter}, J.-M., \& {Racine}, R. 1995, \aj, 109, 1055,
  \dodoi{10.1086/117341}

\bibitem[{{Pooley} {et~al.}(2003){Pooley}, {Lewin}, {Anderson}, {Baumgardt},
  {Filippenko}, {Gaensler}, {Homer}, {Hut}, {Kaspi}, {Makino}, {Margon},
  {McMillan}, {Portegies Zwart}, {van der Klis}, \& {Verbunt}}]{Pooley03}
{Pooley}, D., {Lewin}, W.~H.~G., {Anderson}, S.~F., {et~al.} 2003, \apjl, 591,
  L131, \dodoi{10.1086/377074}

\bibitem[{{Rhode} \& {Zepf}(2001)}]{Rhode2001}
{Rhode}, K.~L., \& {Zepf}, S.~E. 2001, \aj, 121, 210, \dodoi{10.1086/318039}

\bibitem[{{Sahu}(2021)}]{2021wfcd.book....5S}
{Sahu}, K. 2021, in WFC3 Data Handbook v. 5, Vol.~5, 5

\bibitem[{{Sivakoff} {et~al.}(2007){Sivakoff}, {Jord{\'a}n}, {Sarazin},
  {Blakeslee}, {C{\^o}t{\'e}}, {Ferrarese}, {Juett}, {Mei}, \&
  {Peng}}]{Sivakoff07}
{Sivakoff}, G.~R., {Jord{\'a}n}, A., {Sarazin}, C.~L., {et~al.} 2007, \apj,
  660, 1246, \dodoi{10.1086/513094}

\bibitem[{{Spitler} {et~al.}(2016){Spitler}, {Scholz}, {Hessels}, {Bogdanov},
  {Brazier}, {Camilo}, {Chatterjee}, {Cordes}, {Crawford}, {Deneva}, {Ferdman},
  {Freire}, {Kaspi}, {Lazarus}, {Lynch}, {Madsen}, {McLaughlin}, {Patel},
  {Ransom}, {Seymour}, {Stairs}, {Stappers}, {van Leeuwen}, \&
  {Zhu}}]{Spitler2016}
{Spitler}, L.~G., {Scholz}, P., {Hessels}, J.~W.~T., {et~al.} 2016, \nat, 531,
  202, \dodoi{10.1038/nature17168}

\bibitem[{{Sridhar} \& {Metzger}(2022)}]{2022ApJ...937....5S}
{Sridhar}, N., \& {Metzger}, B.~D. 2022, \apj, 937, 5,
  \dodoi{10.3847/1538-4357/ac8a4a}

\bibitem[{{Strader} {et~al.}(2011){Strader}, {Romanowsky}, {Brodie}, {Spitler},
  {Beasley}, {Arnold}, {Tamura}, {Sharples}, \& {Arimoto}}]{Strader2011}
{Strader}, J., {Romanowsky}, A.~J., {Brodie}, J.~P., {et~al.} 2011, The
  Astrophysical Journal Supplement Series, 197, 33,
  \dodoi{10.1088/0067-0049/197/2/33}

\bibitem[{{Tendulkar} {et~al.}(2021){Tendulkar}, {Gil de Paz}, {Kirichenko},
  {Hessels}, {Bhardwaj}, {{\'A}vila}, {Bassa}, {Chawla}, {Fonseca}, {Kaspi},
  {Keimpema}, {Kirsten}, {Lazio}, {Marcote}, {Masui}, {Nimmo}, {Paragi},
  {Rahman}, {Pay{\'a}}, {Scholz}, \& {Stairs}}]{Tendulkar2021NoSF}
{Tendulkar}, S.~P., {Gil de Paz}, A., {Kirichenko}, A.~Y., {et~al.} 2021,
  \apjl, 908, L12, \dodoi{10.3847/2041-8213/abdb38}

\bibitem[{{Thornton} {et~al.}(2013){Thornton}, {Stappers}, {Bailes},
  {Barsdell}, {Bates}, {Bhat}, {Burgay}, {Burke-Spolaor}, {Champion}, {Coster},
  {D'Amico}, {Jameson}, {Johnston}, {Keith}, {Kramer}, {Levin}, {Milia}, {Ng},
  {Possenti}, \& {van Straten}}]{Thornton2013}
{Thornton}, D., {Stappers}, B., {Bailes}, M., {et~al.} 2013, Science, 341, 53,
  \dodoi{10.1126/science.1236789}

\bibitem[{{Tody}(1986)}]{1986SPIE..627..733T}
{Tody}, D. 1986, in Society of Photo-Optical Instrumentation Engineers (SPIE)
  Conference Series, Vol. 627, Instrumentation in astronomy VI, ed. D.~L.
  {Crawford}, 733, \dodoi{10.1117/12.968154}

\bibitem[{{Tody}(1993)}]{1993ASPC...52..173T}
{Tody}, D. 1993, in Astronomical Society of the Pacific Conference Series,
  Vol.~52, Astronomical Data Analysis Software and Systems II, ed. R.~J.
  {Hanisch}, R.~J.~V. {Brissenden}, \& J.~{Barnes}, 173

\bibitem[{{Tolstoy} {et~al.}(2009){Tolstoy}, {Hill}, \& {Tosi}}]{Tolstoy2009}
{Tolstoy}, E., {Hill}, V., \& {Tosi}, M. 2009, \araa, 47, 371,
  \dodoi{10.1146/annurev-astro-082708-101650}

\bibitem[{{Totani}(2013)}]{Totani2013}
{Totani}, T. 2013, \pasj, 65, L12, \dodoi{10.1093/pasj/65.5.L12}

\bibitem[{{Verbunt} \& {Hut}(1987)}]{Verbunt87}
{Verbunt}, F., \& {Hut}, P. 1987, in IAU Symp. 125: The Origin and Evolution of
  Neutron Stars, 187

\bibitem[{{Wang} {et~al.}(2016){Wang}, {Yang}, {Wu}, {Dai}, \&
  {Wang}}]{Wang2016Inspiral}
{Wang}, J.-S., {Yang}, Y.-P., {Wu}, X.-F., {Dai}, Z.-G., \& {Wang}, F.-Y. 2016,
  \apjl, 822, L7, \dodoi{10.3847/2041-8205/822/1/L7}

\bibitem[{{Wenger} {et~al.}(2000){Wenger}, {Ochsenbein}, {Egret}, {Dubois},
  {Bonnarel}, {Borde}, {Genova}, {Jasniewicz}, {Lalo{\"e}}, {Lesteven}, \&
  {Monier}}]{Wenger2000}
{Wenger}, M., {Ochsenbein}, F., {Egret}, D., {et~al.} 2000, \aaps, 143, 9,
  \dodoi{10.1051/aas:2000332}

\bibitem[{{Whitmore} {et~al.}(1993){Whitmore}, {Schweizer}, {Leitherer},
  {Borne}, \& {Robert}}]{Whitmore1993}
{Whitmore}, B.~C., {Schweizer}, F., {Leitherer}, C., {Borne}, K., \& {Robert},
  C. 1993, The Astronomical Journal, 106, 1354, \dodoi{10.1086/116732}

\bibitem[{{Ye} {et~al.}(2023){Ye}, {Fragione}, \&
  {Perna}}]{2023ApJ...953..141Y}
{Ye}, C.~S., {Fragione}, G., \& {Perna}, R. 2023, \apj, 953, 141,
  \dodoi{10.3847/1538-4357/ace1eb}

\end{thebibliography}
\bibliographystyle{aasjournal}

\end{document}